\documentclass[aps,prl,twocolumn,superscriptaddress,showpacs,floatfix,nofootinbib]{revtex4}
\usepackage{amsmath,graphicx}
\begin{document}
\preprint{CERN-PH-TH-2005-???, SPhT-t05/?}

\title{Momentum spectra, anisotropic flow, and ideal fluids}

\author{Nicolas Borghini}
\affiliation{Physics Department, Theory Division, CERN, CH-1211 Geneva 23, 
  Switzerland}
\author{Jean-Yves Ollitrault}
\affiliation{Service de Physique Th\'eorique, CEA/DSM/SPhT, 
  Unit\'e de recherche associ\'ee au CNRS, F-91191 Gif-sur-Yvette Cedex, 
  France.}

\date{\today}

\begin{abstract}
If the matter produced in ultrarelativistic heavy-ion collisions 
reaches thermal equilibrium, its subsequent evolution follows the 
laws of ideal fluid dynamics. We show that general predictions 
can be made on this basis alone, irrespective of the details 
of the hydrodynamical model. 
We derive several scaling rules for momentum spectra and  
anisotropic flow (in particular the elliptic flow, $v_2$, and the 
hexadecupole flow, $v_4$)
of identified particles. Comparison with existing data is briefly 
discussed, and qualitative predictions are made for LHC. 
\end{abstract}


\maketitle

An ultrarelativistic Au-Au collision at RHIC produces 
a dense system of interacting particles and fields, which 
then expands into the vacuum. 
Ideal-fluid models have been succesful in describing 
this expansion~\cite{Kolb:2003dz}: 
they are able, to a certain extent, to reproduce the 
magnitude and the transverse momentum ($p_t$) 
dependence of elliptic flow of identified particles, together with 
their $p_t$ spectra, for momenta $p_t \lesssim 2$\,GeV/c. 

In this Letter, we derive general properties of momentum 
spectra of identified particles emitted by an ideal fluid,
which do not depend on the specific model used. 
We shall introduce an important distinction between slow particles, 
whose velocity equals the fluid velocity at some point, 
and fast particles, whose velocity exceeds the maximum fluid velocity. 
We discuss in detail the implications of ideal-fluid behaviour for
both slow and fast particles.  

The expansion of a plasma into a vacuum goes through successive steps.
If the particle mean free path inside the plasma is small enough, 
the plasma thermalizes, leading to a collision-dominated, isentropic 
expansion, which is described by ideal (i.e., inviscid) fluid dynamics. 
On the contrary, the late stage of the expansion is collision-free. 
In between, a transition regime takes place. This transition regime
must in principle be modelled by transport theory, which encompasses both 
isentropic and  collision-free limits~\cite{Bass:2000ib,Teaney:2001av}.

In the context of heavy-ion collisions, however, the transition regime
is most often modelled by a simple Ansatz, the sudden freeze-out 
approximation, which is a sharp transition between the two extremes:
one first defines a space-time hypersurface $\Sigma$ along the history
of the ideal fluid, on which the transition is expected to take place. 
At each point of $\Sigma$, free-streaming particles are emitted
according to thermal distributions in the rest frame of the fluid. 
Integrating over $\Sigma$, one obtains for 
a given particle the following momentum spectrum~\cite{Cooper:1974mv}:
\begin{equation}
\label{dndp}
E\frac{dN}{d^3{\bf p}} = C\!\int_{\Sigma}
\exp\!\left(-\frac{p^\mu u_\mu(x)}{T}\right) p^\mu d\sigma_\mu,
\end{equation}
where $u^\mu(x)$ is the fluid 4-velocity at point $x$ on $\Sigma$, 
$C$ is a normalization constant, and we have neglected the effects 
of quantum statistics (in practice, the latter may
only be significant for pions at low $p_t$). 
For simplicity, we also assume that the fluid temperature $T$ is 
everywhere the same on $\Sigma$, but the results derived in this paper
do not rely on this assumption. 
The possibility has also been raised that particles of different 
types~\cite{Grassi:1994nf} or with different transverse 
momenta $p_t$~\cite{Tomasik:2002qt} have different freeze-out temperatures. 

Let us briefly discuss the validity of the Cooper-Frye Ansatz, 
Eq.~(\ref{dndp}). 
This approximation is expected to be poor when applied to observables 
which are sensitive to the detailed physics at freeze-out, such as HBT 
radii~\cite{Wiedemann:1999qn}. The reason is that Eq.~(\ref{dndp}) 
assumes an isotropic momentum distribution in the rest frame of the
fluid, while non-relativistic studies have shown that freeze-out 
precisely occurs when the relative difference between parallel and 
transverse components of the kinetic temperature becomes large~\cite{hamel}.
On the other hand, the simple freeze-out Ansatz may be a reasonable
one for computing single-particle spectra, provided that collective
expansion dominates over random, thermal motion. 
This is known as the hypersonic approximation in the context 
of non-relativistic gas dynamics~\cite{hamel,cercignani}. 
In the language of heavy-ion collisions, it can be rephrased as follows:
consistency of the ideal-fluid picture requires that the 
freeze-out temperature be much smaller than the inverse slope 
parameters obtained by exponential fits to transverse-mass 
spectra~\cite{Bearden:1996dd}. 
The best-fit value of the freeze-out temperature at RHIC is 
$T\sim 100$~MeV~\cite{Retiere:2003kf,Hirano:2004rs}, while the inverse
slope parameter for pions is $T_{\rm eff}=210$~MeV~\cite{Adler:2003cb}. 
The value of $T$ is large enough to expect significant deviations from 
ideal-fluid behaviour, i.e., viscous effects~\cite{Teaney:2003pb}. 
It is however interesting to study the small-$T$ limit in 
view of upcoming heavy-ion experiments at LHC, and also 
to have a better grasp on viscous effects, which are easily 
seen as deviations from this limit. 

We therefore investigate systematically the 
properties of momentum spectra in the limit of small $T$. 
The general idea is that the integral over $\Sigma$ in 
Eq.~(\ref{dndp}) can then be performed by means of a saddle-point 
integration. 
\footnote{The same method 
was used earlier to predict the $1/\sqrt{m_t}$ behaviour of 
longitudinal HBT radii~\cite{Makhlin:1987gm}.}
In physical terms, it means that the dominant contribution 
comes from the points where the energy of the particle in the fluid 
frame, $p^\mu u_\mu$, is minimum. 
For a given $p^\mu$, $p^\mu u_\mu$ is a function of the space
components of the fluid 4-velocity, ${\bf u}$ (the fourth one
being related to them through $u^0=\sqrt{1+{\bf u}^2}$), which 
themselves depend on the point $x$ on $\Sigma$. 
Since $p^\mu u_\mu$ is the energy of the particle in the fluid rest 
frame, its absolute minimum is the particle mass $m$. 
This minimum is reached when the particle is at rest with respect 
to the fluid, i.e., when its velocity ${\bf p}/p^0$  
equals the fluid velocity ${\bf u}/u^0$ (or, equivalently, 
if ${\bf u}={\bf p}/m$).  

\begin{figure}[t!]
\centerline{\includegraphics*[width=0.6\linewidth]{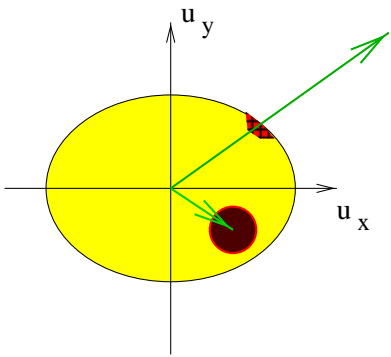}}
\caption{(Color online) 
  Schematic representation of the distribution of the fluid transverse 
  4-velocity. The $x$-axis is the impact-parameter direction. 
  The short (resp. long) arrow indicates the four-velocity of a slow 
  (resp. fast) particle. The most probable values of the fluid four-velocity 
  for this particle are marked as darker (red) areas.}
\label{fig:fig1}
\end{figure}

This absolute minimum, however, occurs only if there exists a point 
on $\Sigma$ where this value of the fluid velocity is reached. 
This leads us to a qualitative discussion of the values taken by 
${\bf u}$ at freeze-out. The longitudinal fluid velocity is 
expected to span almost the whole range from $-1$ to $1$ 
in ultrarelativistic collisions, and the simple Bjorken 
picture $u^z/u^0=z/t$~\cite{Bjorken:1982qr} shows, at least qualitatively, 
how it is related to space-time coordinates. 
The radial fluid velocity spans a more limited range. 
The reason is that 
transverse collective flow is not initially present in the system
but builds up progressively. For a given fluid rapidity 
$y_f=\frac{1}{2}\ln((u^0+u^z)/(u^0-u^z))$, the transverse 4-velocity, 
$\sqrt{u_x^2+u_y^2}$, extends up to some maximum value $u_{\rm max}$, 
which may depend on $y_f$, and on the azimuthal angle $\phi_f$ 
for non-central collisions. 
As shown schematically in Fig.~\ref{fig:fig1}, $u_{\rm max}(y_f,\phi_f)$ 
is largest at $\phi_f=0$, along the direction of impact parameter. 
This is due to larger pressure 
gradients in this direction~\cite{Ollitrault:1992bk}, 
which explain the large in-plane elliptic flow observed at 
RHIC~\cite{Ackermann:2000tr}. 
Typical values of $u_{\rm max}$ are of order 1 at 
RHIC ($u_{\rm max}=\sinh\rho_0$ in the notations
of~\cite{Retiere:2003kf}, with the best-fit value $\rho_0\simeq
0.9$). 

{}From now on, we make a distinction between ``slow'' and ``fast'' particles 
as follows: a particle of mass $m$, with rapidity $y$ and transverse 
momentum $p_t$, is defined as slow if $p_t/m<u_{\rm max}(y,\phi)$ for 
all $\phi$ (that is, actually, for $\phi=\pi/2$, where the minimum occurs). 
Conversely, a fast particle is defined by  $p_t/m>u_{\rm max}(y,\phi)$ for 
all $\phi$ (that is, for $\phi=0$, where the velocity is maximal). 
Between both regimes, there is a small intermediate 
region, which will not be considered in this Letter. 

For a slow particle, there is a point on $\Sigma$ such 
that the fluid velocity equals the particle velocity, and 
the minimum $p^\mu u_\mu=m$ is reached. (Our results are unchanged 
if there are two or more such points.)
If $T$ is small enough, the dominant contribution to the 
integral (\ref{dndp})  comes from the neighbourhood of this
point (see Fig.~\ref{fig:fig1}). 
The integral can then be evaluated approximately by expanding 
the exponent to second order around the minimum of $p^\mu u_\mu$. 
What remains is a Gaussian integral. 
For a given velocity, $p^\mu$ is proportional to the particle mass
$m$. Hence, the width of the Gaussian varies with $m$ like 
$1/\sqrt{m}$, and the integral over $\Sigma$ in Eq.~(\ref{dndp})
is $m^{-3/2}$ times a function of the particle velocity. 
This means that the mass dependence is only a global factor 
for slow particles:
\begin{equation}
\label{spectraslow}
E\frac{dN}{d^3{\bf p}} \equiv \frac{dN}{p_t\,dp_t\, d\phi\, dy} =
c(m) f\!\left(\frac{p_t}{m},y,\phi\right), 
\end{equation}
where $f$ is the same for all particles. 
As a result, transverse momentum and rapidity spectra 
(integrated over $\phi$) of identified slow particles 
coincide, up to a normalization factor, when they are plotted
as a function of $p_t/m$ and $y$. 
The coefficients quantifying azimuthal anisotropies 
$v_n=\langle\cos n\phi\rangle$, which are independent of the total 
yield, should coincide for different identified slow particles 
at the same $p_t/m$ and $y$. 
This property holds for directed flow $v_1$ as well as for elliptic flow
$v_2$ and the higher harmonics. 
The behaviour is qualitatively correct for $v_2$ at RHIC, which rises 
more slowly with $p_t$ for heavier 
particles~\cite{Adams:2004bi,Adler:2003kt}. 

The condition under which the saddle-point approximation 
is good for slow particles can be roughly stated 
as $T\ll m$ for a relativistic fluid, for dimensional reasons: 
the larger $m$, the smaller 
the width of the Gaussian, and the better the approximation. 
A detailed calculation gives the condition 
\begin{equation}
\label{conditionslow}
T\ll mv_{\rm max}^2
\end{equation} 
with $v_{\rm max}\equiv u_{\rm max}/u^0_{\rm max}$ and 
$u^0_{\rm max}=\sqrt{1+u_{\rm max}^2}$; this condition amounts 
to assuming that collective motion 
dominates over thermal motion.
At RHIC, we expect the approximation to be poor for pions 
(furthermore, the pion spectrum is 
contaminated at low $p_t$ by secondary decays, and may also be sensitive 
to Bose--Einstein statistics), but it might be a reasonable one 
for kaons and heavier hadrons with $p_t<m$/c. 

Let us now discuss fast particles. 
Roughly speaking, these are the particles that move faster 
than the fluid: the minimum value of $p^\mu u_\mu$  is larger
than $m$. In order to locate this minimum, we denote by 
$y$ (resp. $y_f$) the particle (resp. fluid) longitudinal rapidity, by 
$u_\parallel$ the transverse component of ${\bf u}$  parallel to the particle
transverse momentum ${\bf p}_t$, and by $u_\perp$ the transverse
component of ${\bf u}$ orthogonal to ${\bf p}_t$. 
With these notations, 
\begin{equation}
p^\mu u_\mu = m_t\cosh(y-y_f)\sqrt{1+u_\parallel^2+u_\perp^2}
-p_t u_\parallel,
\end{equation}
where $m_t\equiv\sqrt{m^2+p_t^2}$.
Minimization with respect to $y_f$ and $u_\perp$ gives 
$y_f=y$ and $u_\perp=0$, i.e., 
the fluid velocity is {\it parallel\/} to the particle
velocity. 
The minimum is then attained when $u_\parallel$ is maximum, i.e, 
when $u_\parallel=u_{\rm max}(y,\phi)$, where $\phi$ is the 
azimuthal angle of the particle. 
In other words, fast particles come from regions on $\Sigma$ where 
the parallel velocity is close to its maximum value (see 
Fig.~\ref{fig:fig1}). 
A saddle-point integration then gives~\footnote{We assume that the 
maximum value $u_{\rm max}$ is reached at an inner point 
of $\Sigma$. If it occurs at the edge of the fluid, there is no 
square root in the pre-exponential factor.}
\begin{equation}
\label{dndpfast}
\frac{dN}{dy\, d^2{\bf p}_t} \propto
\frac{1}{\sqrt{p_t-m_t v_{\rm max}}}
\exp\!\left(\frac{p_t u_{\rm max}-m_t u^0_{\rm max}}{T}\right),
\end{equation}
where the $(y,\phi)$ dependence is implicit.
This result was already obtained long ago for massless 
particles in Ref.~\cite{Blaizot:1986bh} 
(see also \cite{Schnedermann:1993ws}). 

The saddle-point approximation 
is valid for fast particles if $p_t$ is large enough. 
A more precise criterion is
\begin{equation}
\label{condition}
T\ll \frac{(p_t u^0_{\rm max}-m_t u_{\rm max})^2}
{m_t u^0_{\rm max}-p_t u_{\rm max}},
\end{equation}
together with the condition $p_t>m u_{\rm max}$. 
At RHIC, the left-hand side (lhs) of Eq.~(\ref{condition}) is smaller 
than the right-hand side (rhs) by at least a factor of 2 as soon as 
$p_t>0.7$~GeV/c for pions, $p_t>1.2$~GeV/c for kaons, 
and $p_t>1.8$~GeV/c for (anti)protons. 
On the other hand, ideal fluid dynamics is expected to 
break down if $p_t$ is too high, since high $p_t$ particles 
have been shown to be more sensitive to off-equilibrium (viscosity) 
effects~\cite{Teaney:2003pb}. Deviations from fluid-like behaviour
are best seen on elliptic flow, for mesons above 1.5~GeV/c, and 
for baryons above 2.5~GeV/c. 
The window in which our approximation works is likely to be 
narrow, which reflects the importance of viscous effects at RHIC. 
Better agreement should be reached at LHC. 

The $p_t$ spectra of identified particles 
are directly obtained from Eq.~(\ref{dndpfast}), 
neglecting the $\phi$ dependence of $u_{\max}$.
Radial flow results in flatter $m_t$-spectra for 
heavier particles. In addition, Eq.~(\ref{dndpfast}) 
implies a breakdown of $m_t$-scaling: the slope of 
the spectrum decreases with increasing $m_t$ for pions, 
and increases for protons, in qualitative agreement with 
experimental findings~\cite{Adler:2003cb}.

For non-central collisions, we can also obtain the anisotropic flow 
coefficients. 
We expand $u_{\rm max}(\phi)$ in Fourier series, and neglect 
odd harmonics:
\begin{equation}
\label{defV2V4}
u_{\rm max}(\phi)=u_{\rm max}(1+2 V_2\cos(2\phi)+2 V_4\cos(4\phi)+\cdots).
\end{equation}
The parameter $V_2$ is of the order of 4~\% for semi-central Au-Au
collisions at RHIC. It is related to the parameter $\rho_2$ of  
blast wave parameterizations~\cite{Huovinen:2001cy,Retiere:2003kf}
by $V_2=\rho_2/(2 v_{\rm max})$.  
The $\phi$ distribution is obtained by inserting 
Eq.~(\ref{defV2V4}) into (\ref{dndpfast}). 
If $T$ is small enough, the $\phi$  
dependence in Eq.~(\ref{dndpfast}) is dominated by the 
exponential. 
Expanding the latter to first order in $V_2$, one obtains 
\begin{equation}
\label{v2fast}
v_2(p_t)=\frac{V_2 u_{\rm max}}{T}\left(p_t-m_t v_{\rm max}\right).
\end{equation}
A similar equation was already obtained in Ref.~\cite{Huovinen:2001cy} 
in the framework of a simplified fluid model, and was shown to fit 
RHIC data rather well. 
In particular, Eq.~(\ref{v2fast}) shows that the ``mass ordering'' 
which follows from Eq.~(\ref{spectraslow}) for slow particles 
persists at high $p_t$ in hydro: 
at a given $p_t$, heavier particles have smaller $v_2$. 

We finally make predictions for the hexadecupole flow, $v_4$. 
We expand the exponential of Eq.~(\ref{dndpfast}) and look for 
terms in $\cos(4\phi)$. To leading order one obtains two terms:
\begin{eqnarray}
\label{v4fast}
v_4(p_t) &\!=\!& 
\frac{(V_2 u_{\rm max})^2}{2T^2}\left(p_t-m_t v_{\rm max}\right)^2 \cr 
 & & +\,\frac{V_4 u_{\rm max}}{T}\left(p_t-m_t v_{\rm max}\right).
\end{eqnarray}
For large enough $p_t$, the first term dominates over the second, 
which gives the simple, universal relation 
\begin{equation}
\label{v4fast2}
v_4(p_t)=\frac{v_2(p_t)^2}{2}.
\end{equation}
Let us derive the domain of validity of this approximation. 
If $u_{\rm max}(\phi)$ is a smooth function of $\phi$, 
one generally expects $V_4$ to be of order $(V_2)^2$. 
The condition for Eq.~(\ref{v4fast2}) is then 
\begin{equation}
\label{condition2}
T\ll u_{\rm max}\left(p_t-m_t v_{\rm max}\right).
\end{equation}
At RHIC, the lhs is smaller than the rhs by at least a factor of 2
for pions with $p_t>0.8$~GeV/c. For heavier particles, 
Eq.~(\ref{condition}) supersedes  Eq.~(\ref{condition2}).

Our result, Eq.~(\ref{v4fast}), is in contradiction with the statement 
that $v_4$ is 
a sensitive probe of initial conditions~\cite{Kolb:2003zi}: on the contrary, 
we find a universal result, which can be directly used as a probe
of {\em ideal fluid behaviour\/}, not of initial conditions. 
The experimental value found by the STAR 
Collaboration~\cite{Adams:2003zg}
is a factor of 2 to 3 higher than our prediction.
Deviations from ideal-fluid behaviour are generally 
expected to yield higher values of $v_4$~\cite{Bhalerao:2005mm}.

\begin{figure}[t!]
\centerline{\includegraphics*[width=\linewidth]{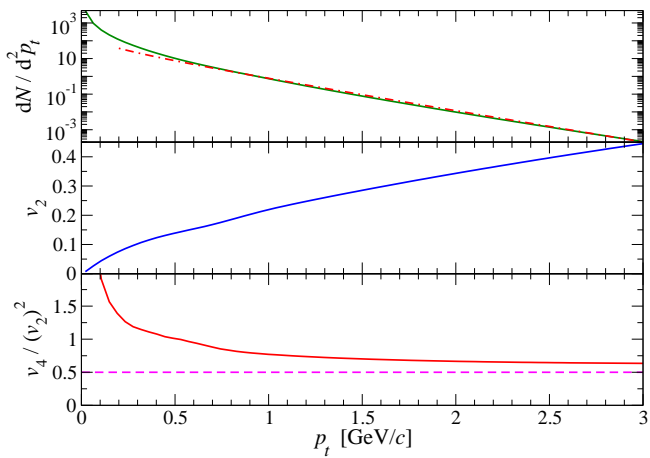}}
\caption{(color online) 
Numerical results from ideal hydrodynamics with 
Bjorken longitudinal expansion~\cite{Bjorken:1982qr}.
and a black-body equation of state. 
Initial conditions mimic a midcentral ($b$=8~fm) 
Au-Au collision at RHIC. The initial density has been 
fixed to reproduce $\langle p_t\rangle=450$~MeV/c, as measured
for pions at RHIC~\cite{Adler:2003cb}. 
Top: Solid line: $dN/dyd^2 p_t$; dash-dotted line, fit using 
Eq.~(\ref{dndpfast}). Middle: $v_2(p_t)$. 
Bottom: $v_4(p_t)/v_2(p_t)^2$. }
\label{fig:fig2}
\end{figure}

Our results for fast particles, 
Eqs.~(\ref{dndpfast}), (\ref{v2fast}), (\ref{v4fast2}), are
compared to results from a numerical 3-d hydrodynamical calculation in 
Fig.~\ref{fig:fig2}. 
The calculation has been pushed to very large times, 
so that the small-$T$ limit applies. 
The value of $v_4/v_2^2$ does not go exactly to 0.5 at large $p_t$ 
but rather to 0.63. This is due to the fact that 
the initial eccentricity is large for this value of the impact
parameter, and Eq.~(\ref{v4fast2}) is obtained through a leading
order expansion in the anisotropy. We have checked numerically 
that agreement is better for lower values of $b$, where the
eccentricity is smaller. 

Before we come to our conclusions, let us compare our 
approach with the popular blast-wave one. The 
blast-wave parameterization, in its simplest form, assumes a unique radial 
velocity for the fluid~\cite{Siemens:1978pb}; this framework has 
recently been refined to take into account the azimuthal dependence 
of the fluid velocity~\cite{Huovinen:2001cy} and of the freeze-out
surface~\cite{Adler:2001nb} in non-central collisions, and even a 
distribution of fluid velocities~\cite{Retiere:2003kf}. 
A few parameters (typically four) are then fitted to experimental 
data. 
Some of the results we derived above were already obtained within the 
blast-wave approach, namely the mass-ordering of the $v_2(p_t)$ of 
different types. 
However, our present framework is more general in the sense that we do not 
assume a given fluid-velocity profile, but also more specific in the sense
that we assume that collective motion dominates over thermal (random)
motion. 
In addition, blast-wave fits treat slow and fast particles on an equal 
footing, ignoring the distinction between both types of particles. 
Although fitting the whole spectrum with a single formula is admittedly 
more convenient, it misses an important feature of the underlying physics, 
since slow and fast particles originate from different regions of the 
expanding fluid. 
In particular, fits using our formulas for fast particles may yield 
values of $T$ and $u_{\rm max}$ which differ from blast-wave fits. 
Finally, our formulas are significantly simpler than 
blast-wave parameterizations, which involve special functions.

We have obtained the following results for momentum 
spectra and anisotropies in the framework of ideal-fluid models
using a saddle-point approximation of the momentum distribution:
\begin{itemize}
\item At low $p_t$, 
identified particles of different masses have 
the same momentum spectra and anisotropies (up to a normalization
for the spectra), when plotted as a function of velocity 
variables $y$ and $p_t/m$. 
This defines ``slow'' particles. 
This scaling is due to the fact that slow particles move with 
the fluid: they come from the regions where the fluid velocity 
equals their velocity. 
The scaling is expected to be poor for pions. It is expected to break 
down when $p_t/m$ exceeds $u_{\rm max}$, the maximum 
value of the transverse 4-velocity of the fluid. 
$u_{\rm max}$ may in general depend on the rapidity $y$, and 
reflects the underlying equation of state of the expanding matter. 
\item 
Fast particles, defined by Eq.~(\ref{condition}), 
all originate from the region where the fluid is fastest along
the direction of the particle velocity. As a 
result, their transverse momentum spectra and azimuthal anisotropies 
at a given rapidity are uniquely determined by three 
parameters $u_{\rm max}$, $T$, and $V_2$, and given by 
Eqs.~(\ref{dndpfast}), (\ref{v2fast}), (\ref{v4fast2}).
Comparing the $v_2$ of different particles should directly 
give the precise value of $u_{\rm max}$, while transverse 
momentum spectra yield $T$. 
\end{itemize}
These results can be used as signatures of hydrodynamic evolution 
in heavy-ion collisions, and also as consistency checks of 
numerical ideal-fluid calculations. 
Ideal-fluid evolution leads to different 
behaviours for slow and fast particles. 
Some of the results obtained for fast particles (in particular 
for elliptic flow) are already known from blast-wave approaches.
We have shown that they are in fact more general.
The scaling rules for slow particles, which are evidenced here for 
the first time, should be further tested on available RHIC data. 
We expect all our results to be in closer agreement with data 
at LHC than at RHIC. In particular, we predict that the value 
of the ratio  $v_4/(v_2)^2$ should be lower at LHC than at RHIC.

\section*{Acknowledgments}

J.-Y. O. thanks F. Becattini, T. Hirano, 
E. Shuryak and R. Snellings for discussions. 
We thank J.-P. Blaizot for careful reading of the manuscript, 
and the referee for pointing out a mistake in the derivation of 
Eq.~(\ref{v4fast2}).

\end{document}